\documentclass[final,5p,times,twocolumn]{elsarticle}
\usepackage{lineno,hyperref}
\usepackage{amssymb}
\usepackage{amsmath}
\usepackage{mwe}
\usepackage{siunitx}
\usepackage{tabularx}
\usepackage{algorithmic}
\usepackage{caption}
\usepackage{algorithm2e} 
\usepackage{amsthm}
\usepackage{graphicx}
\usepackage{lscape}
\usepackage{verbatim}
\usepackage{color,soul}
\usepackage{xcolor}
\usepackage{subfigure}
\usepackage{tabularx,ragged2e,booktabs,caption,array,multirow,multicol}
\usepackage{csquotes}
\usepackage{mathtools}
\usepackage{lineno}
\usepackage{amsthm}  
\usepackage{listings}
\usepackage{amssymb}
\usepackage{latexsym}
\usepackage{epsfig}
\usepackage{float}
\usepackage{xspace} 
\usepackage{wrapfig}

\usepackage{tikz}
\usetikzlibrary{shapes,arrows}

 \newcommand{\mb}[1]{\mathbf{#1}}

\modulolinenumbers[5]

\journal{Computational Geoscience}

\begin{document}

\begin{frontmatter}

\title{ Surrogate-assisted distributed swarm optimisation for computationally  expensive geoscientific  models}

\author[geo]{Rohitash Chandra \corref{corauthor}  }
\ead{rohitash.chandra@unsw.edu.au} 

\author[iit]{Yash Vardhan Sharma}

\address[geo]{Transitional Artificial Intelligence Research Group, School of Mathematics and Statistics, University of New South Wales, Sydney, Australia}

\address[iit]{Mechanical and Industrial Engineering Department, Indian Institute of Technology Roorkee, India}

\cortext[corauthor]{Corresponding author}
\cortext[contrib]{Authors contributed equally}

\newcommand\eindent{\endgroup} 
 \newcommand{\bs}[1]{\boldsymbol{#1}}
 \newcommand{\mc}[1]{\mathcal{#1}}%

\begin{abstract}  

Evolutionary algorithms provide gradient-free optimisation which is beneficial for models that have difficulty in obtaining gradients; for instance, geoscientific landscape evolution models. However, such models are at times computationally expensive and even distributed swarm-based optimisation with parallel computing struggles. We can incorporate efficient strategies such as surrogate-assisted optimisation to address the challenges; however, implementing inter-process communication for surrogate-based model training is difficult.    In this paper, we implement surrogate-based estimation of fitness evaluation in distributed swarm optimisation over a parallel computing architecture. \textcolor{black}{We first test the framework on a set of benchmark optimisation problems and then apply it to a geoscientific model that features a landscape evolution model.} Our results demonstrate very promising results for benchmark functions and the Badlands landscape evolution model. We obtain a reduction in computational time while retaining optimisation solution accuracy through the use of surrogates in a parallel computing environment.  The major contribution of the paper is in the application of surrogate-based optimisation for geoscientific models which can in the future help in a better understanding of paleoclimate and geomorphology.

\end{abstract}


\end{frontmatter}

\section{Introduction}

Evolutionary algorithms are loosely motivated by the theory of evolution where species are represented  by individuals in a population that compete and collaborate with each other, producing offspring over generations that to improve quality given by fitness measure  \citep{davis1991handbook,LI202010}. \textit{Particle swarm optimisation} (PSO), on the other hand, is motivated by the flocking behaviour of birds or swarms    represented by a population of particles  (individuals) that compete and collaborate over time \citep{ab2015comprehensive,fang2010review,freitas2020particle,jain2022overview}. Evolutionary and swarm optimisation methods have been prominent in a number of areas such as real-parameter global optimization,  
combinatorial optimization and scheduling, and   machine learning   \citep{man1996genetic,ponsich2012survey,moriarty1999evolutionary,ab2015comprehensive}. Research in evolutionary and swarm optimisation has focused on different  ways to create new solutions with mechanisms that are heuristic in nature;  hence, different variants are available \citep{larranaga1999genetic,fang2010review}.  A major challenge has been in applying them in large-scale or computationally expensive optimisation problems that require thousands of function evaluations where a single function evaluation can take minutes  to hours, or even days \citep{holden2018abc}. An example of an expensive function is a geoscientific model for landscape evolution problem \citep{chandra2020surrogate}, and  deep learning models for  big data problems \citep{zhang2018survey}.  Computationally expensive optimisation can be addressed with distributed/parallel computing   and swarm optimisation \citep{bakwad2011fast,sivanandam2009dynamic,alba2002parallelism}; however, we need efficient strategies for representing the problem.
 
 
  \textit{Surrogate-assisted optimisation} \citep{hicks1978wing,jin2011surrogate, ong2003evolutionary,zhou2007combining,he2023review} provides a remedy for expensive models with  the use of statistical and machine learning models to provide a low computational replicate of the actual model. The  surrogate model  is developed by training from available data generated during optimisation that features a set of inputs (new solutions)  and corresponding  output (fitness) given by the actual  model.  The method is also known as  Bayesian optimisation where the surrogate model  (acquisition function) is typically  a Gaussian process model \citep{calandra2016bayesian}; however, neural networks and other machine learning models have also been  used \citep{shahriari2015taking}.    Evolutionary and swarm optimization methods have been used in surrogate-assisted and Bayesian optimization, and have been prominent  in fields of  engine  and aerospace design  \citep{ong2005surrogate,jeong2005efficient,samad2008multiple,hicks1978wing}, robotics \citep{calandra2016bayesian}, experimental design \citep{greenhill2020bayesian},  and machine learning \citep{snoek2012practical,shahriari2015taking}.  
 
 \textcolor{black}{ Although limited studies exist, surrogate-assisted optimisation has been applied to geoscience problems. Wang et al. \cite{wang2023reliability} presented reliability-enhanced surrogate-assisted PSO in landslide displacement prediction where the method was used for feature selection and hyperparameter optimization. The PSO-based  surrogate model  was used to  search the hyperparameters and feature sets in the long short-term memory (LSTM) deep learning model  for predicting landslide displacement.  Gong et al. \cite{gong2022ensemble} presented  an ensemble-based  surrogate-assisted cooperative PSO  for water contamination source identification.  Zhou et al.   \cite{zhou2023hierarchical} used a surrogate-assisted evolutionary algorithm that incorporated multi-objective optimization for  oil and gas reservoirs  focusing on good placement and hydraulic fracture parameters. The method employed global–local hybridization searching strategy via PSO with  low-fidelity surrogate model that used a multilayer perception. Furthermore, Zhand et al. \cite{zhang2022parameter} used surrogate-assisted PSO for  gas hydrate reservoir development. Wang et al. \cite{wang2022adaptive} presented a surrogate-assisted model   for   the optimization  of  hyperspectral remote sensing images.   Chen et al. \cite{chen2022surrogate} utilised a surrogate-assisted evolutionary algorithm  for heat extraction optimization of enhanced geothermal systems. 
    }

     Evolutionary algorithms provide gradient-free optimisation which is beneficial for models   that do not have gradient information, for instance,  landscape evolution models \citep{chandra2019multicore,chandra2020surrogate}. Some instances of such  models are so expensive that even distributed evolutionary algorithms with the power of parallel computing would struggle. Hence, we need to incorporate efficient strategies such as surrogate-assisted optimisation that further improves their performance, but this becomes a challenge given parallel processing and inter-process communication for implementing surrogate estimation.
     
    Landscape evolution models are geoscientific models that can be used to reconstruct the evolution of the Earth's landscape over tens of thousands  millions of years \textcolor{black}{\citep{coulthard2001landscape,chen2014landscape,bishop2007long,martin2004numerical,barnhart2020projections}}. These models guide geologists and climate scientists in better understanding Earth's geologic and climate history that can further also help in foreseeing the distant future of the planet \citep{bishop2007long,temme2009can}. These models use data from geological observations such as bore-hole data and estimated landscape topography millions of years ago and require climate conditions and geological parameters that are not easily known. Hence, it becomes an optimisation problem to estimate these parameters which  has been tackled mostly with Bayesian inference  via Markov Chain Monte Carlo (MCMC) sampling  in previous studies that used parallel computing \citep{chandra2019multicore}, and surrogate assisted Bayesian inference \citep{ChandraGMD2020}. Motivated by these studies, we bring the problem of the landscape evolution model to the evolutionary and swarm optimisation community; rather than viewing it as an inference problem, we view it as an optimisation problem.

 In this paper, we implement a surrogate-based optimisation framework via swarm optimisation over a parallel computing architecture. We apply the  framework  for benchmark optimisation functions and a selected  landscape evolution model. We investigate performance measures such as the accuracy of surrogate prediction given different types of problems that differ in terms of dimension and fitness landscape. \textcolor{black}{The contribution of the paper is in the application of surrogate-based optimisation for geoscientific models using parallel computing. The estimation  of parameters through surrogate-assisted estimation can in the future help in better understanding paleoclimate and geomorphology which can enhance knowledge about climate change.}

 The rest of the paper is organised as follows. Section 2 provides background and related work, while Section 3 presents the proposed  methodology. Section 4 presents experiments and results and Section 5 concludes the paper with a discussion for future research. 
 
 \section{Related work}

 \subsection{Surrogate-assisted optimization}

 PSO has been continuously becoming prominent in surrogate-assisted optimisation. Yu   \emph{et al.} \citep{yu2018surrogate} compared  surrogate-assisted hierarchical PSO, standard  PSO,  and a social learning PSO for selected  benchmark functions   under a limited computational budget. Li  \emph{et al.} \citep{LI2020} presented a  surrogate-assisted PSO for computationally expensive problems where two criteria were applied in tandem to select candidates for exact evaluations. \textcolor{black}{These included a  performance-based criterion   and a distance-based   criterion used to enhance   exploration  that does not consider the fitness landscape of different problems.} The  results demonstrated better performance over several state-of-the-art algorithms for selected benchmark functions and  a propeller design problem. Chen   \emph{et al.} \citep{CHEN2021} presented a hierarchical surrogate-assisted differential evolution  algorithm for  high-dimensional expensive optimization problems with RBF network  for selected benchmark functions  and an oil reservoir production optimization problem. Yi  \emph{et al.} \citep{YI20191} presented an online variable-fidelity surrogate-assisted harmony search algorithm with a multi-level screening strategy that showed promising performance for expensive engineering design optimization problems.

Furthermore,  Li  \emph{et al.} \citep{LI2019291} presented an ensemble of surrogates assisted PSO of medium-scale expensive problems which used multiple trial positions for each particle  and selected the promising positions by using the superiority and uncertainty of the ensemble simultaneously. In order to feature faster convergence and to avoid  wrong global attraction of models, the optima of two surrogates that featured polynomial regression  and RBF models were  evaluated in the convergence state of particles.  Liao  \emph{et al.} \citep{LIAO2020106262}  presented multi-surrogate multi-tasking optimization of expensive problems to accelerate the convergence by regarding the two surrogates as two related tasks. Therefore, two optimal solutions found by the multi-tasking algorithm were evaluated using the real expensive objective function, and  both the global and local models were updated  until the  computational budget was exhausted.   The results indicated  competitive performance with faster convergence that scaled well with an increase in problem dimension for solving computationally expensive single-objective optimization problems. Dong  \emph{et al.} \citep{DONG2020100713} presented surrogate-assisted black wolf optimization for high-dimensional and computationally expensive black-box problems that  featured RBF-assisted meta-heuristic exploration.  The RBF featured  knowledge mining that includes a global search carried out using the black wolf optimization and a local search  strategy combining global and multi-start local exploration. The method obtained superior computation efficiency and robustness demonstrated by comparison tests with benchmark functions. Chen  \emph{et al.} \citep{CHEN2021} presented efficient hierarchical surrogate-assisted differential evolution for high-dimensional expensive optimization using global and local surrogate models  featuring RBF network with an application to an oil reservoir production optimization problem. The results show that the method was effective for most benchmark functions and gave a promising performance for a reservoir production optimization problem. \textcolor{black}{
Ji   \emph{et al.} \citep{ji2021dual} presented a  dual-surrogate-assisted cooperative PSO for expensive multimodal problems  which reported highly competitive optimal solutions at a low computational cost for benchmark problems and a building energy conservation problem.  Ji  \emph{et al.} \citep{ji2021multisurrogate} further extended their previous approach using multi-surrogate-assisted multitasking PSO  for expensive multimodal problems.}
 
  Some of the prominent examples of surrogate-assisted approaches in the Earth sciences include modelling  water resources \citep{razavi2012review,asher2015review}, computational oceanography \citep{van2007fast},  atmospheric general circulation models  \citep{Scher2018}, carbon-dioxide   storage and oil recovery   \citep{ampomah2017co}, and debris flow models \citep{navarro2018surrogate}.  
   
 \subsection{Landscape evolution models and Bayeslands}

  Landscape evolution models (LEMs) use different climate and geophysical  aspects  such as tectonics or climate variability \citep{Whipple2002,Tucker10,salles2018pybadlands,Campforts2017,Adams2017} and  combine empirical data and conceptual methods into a set of mathematical equations that form the basis for driving model simulation. \textit{Badlands} (basin and landscape dynamics) \citep{salles2018pybadlands,salles2016badlands} is a LEM that    simulates landscape evolution and sediment transport/deposition  \citep{Howard1994,Hobley2011} with an initial topography exposed to climate and geological factors over time with given  conditions (parameters) such as the \textit{precipitation} rate  and rock \textit{erodibility} coefficient. The major challenge is in estimating the climate and geological parameters and those that are linked with  the initial topography since they  can range millions of years in geological timescale depending on the problem. A way is to develop an optimisation framework that utilizes limited data. Since gradient information is  not available, Badlands can be seen as a black-box optimisation model where the unknown parameters need to be found via optimisation or Bayesian inference. So far, the problem has been approached with Bayesian inference that employs MCMC sampling for the estimation of these parameters in a framework known as Bayeslands \citep{chandra2019bayeslands}. 
  
  The Bayeslands  framework had limitations due to the computational complexity of the Badlands model; hence, in our earlier works, we extended it using parallel tempering MCMC \citep{chandra2019PT-Bayeslands} that  featured parallel computing to enhance  computational efficiency.  Although we used parallel computing with a small-scale synthetic Badlands model, the procedure remained  computationally challenging since thousands of samples were drawn and evaluated. Running a single large-scale real-world Badlands model can take several minutes to  hours, and even several  days depending on the area covered by the landscape evolution model considered, and the span of geological time considered in terms of millions of years. Therefore,   parallel tempering Bayeslands was further enhanced through surrogate-assisted estimation. We used developed surrogate-assisted parallel tempering MCMC for landscape evolution models where a global-local surrogate framework  utilised surrogate training in the main process that   managed  MCMC replicas running in parallel \citep{ChandraGMD2020}.  We obtained  promising results, where the prediction performance was  maintained while lowering the overall computational time.

  \section{Surrogate-assisted distributed swarm optimisation}

   \subsection{PSO}
    
PSO is a population-based metaheuristic that  improves the population over iterations with  a given measure of accuracy  known as fitness \citep{kennedy1995particle}. The population of the candidate solution   is known as a swarm  while the candidate solutions are known as particles that get updated  according to  the particle's position and velocity. In the swarm,   each particle's movement is typically  influenced by its local best-known position which  gets updated when  better positions are discovered  by other particles. Equations \ref{eq:1} and \ref{eq:2} show the velocity and position update of the particle in a swarm, respectively.
\begin{equation}
\mathbf{v}_{t+1} = \alpha \mathbf{v}_{t} + c_1 \gamma_1 (\mathbf{x}_{pbest} - \mathbf{x}_{t}) + c_2 \gamma_2 (\mathbf{x}_{gbest} - \mathbf{x}_{t})
\label{eq:1}
\end{equation} 
\begin{equation}
    \mathbf{x}_{t+1} = \mathbf{x}_{t} + \mathbf{v}_{t+1}
\label{eq:2}
\end{equation}
 
 where $\boldsymbol{v_t}$  $\boldsymbol{x_t}$ represent the velocity and position of a particle  at time step $t$, respectively.  $c_1$ and $c_2$ are the user-defined cognitive and social acceleration coefficients, respectively.  $\gamma_1$ and $\gamma_2$ are random numbers drawn from uniform $U[0,1]$ distribution, and $\alpha$ is a user-defined inertia weight. There are several variants in the way the particles get updated  which have their strengths and limitations for different types of problems \textcolor{black}{\citep{kennedy1995particle,shi2001particle,van2004cooperative,zhan2009adaptive,yang2004quantum,jain2022overview}}.

   
 \subsection{Surrogate assisted Distributed framework} 
   
We update the swarm using the canonical particle update method \citep{kennedy1995particle} and execute the swarms in a distributed swarm   framework that employs parallel computing. 

\subsubsection{Inter-process Communication}

In our  framework, we  exchange selected   swarm particles after a certain number of generations with inter-process communication. During the exchange, we replace   20 percent of the weaker particles   with stronger ones from other swarms. \textcolor{black}{This enhances the exploration and exploitation properties of our framework  for the optimization problem}. Afterwards, the  process continues where  local swarms create  particles with new positions and velocities as shown in Figure \ref{fig:surrogate}. Hence, we feature distributed swarms and parallel processing for better  diversity and  computational complexity. We execute distinct parallel processes for respective  swarms  with central processing units (CPUs).  Each process  features the optimisation function which is either a synthetic   problem made to be computationally expensive using time delay,   or an application problem. 

\subsubsection{Surrogate Model}

Suppose  that the  true function (model) is represented as $F = g(x)$; where $g()$ is the function and $x$ is a solution or particle from a given swarm. Our  surrogate model outputs pseudo-fitness $\hat{F} = \hat{g}(x) $ that would give  an approximation of the true function via $F = \hat{F} + e$; where,  $e$ represents the difference between the surrogate and true function. 
 The surrogate model   gives an estimate using the \textit{pseudo-fitness}  for replacing  the true function when required by the framework.

 $S_{prob}$ is an important hyperparameter that controls the use of a surrogate in the prediction. We do not want it to be too high in case we are not very confident, i.e. when enough data is not present for our surrogate model. We  do not want it to be too low since  the entire optimisation process will become time-consuming; hence, we need to tune this parameter. The surrogate model is trained by accumulating the data from all the swarms, i.e. the  input $\bf{x}_{i,s}$ and associated  true-fitness $F_{i,s}$ pairs; where $s$ represents the particle  and $i$ represents the swarm. In order to benefit from the surrogate model in the optimization process, it's very crucial to manage the surrogate training and surrogate usage. We cannot use a surrogate from the very beginning of the optimization as its predictions will be random, and we also cannot wait too long as the computational efficiency will be affected. Hence, to manage the surrogate training and its use, an important hyperparameter $\psi$ is used which refers to the surrogate interval measured in terms of the number of generations. $\psi$  is the interval from which  the collected data is used to train the model. The updated surrogate model is used   until the next interval is reached, and knowledge in the surrogate model is refined in the next stage (defined by $\psi$)  -- this can be seen as a form of transfer learning. The collected input features ($\Phi$) combined with the true fitness $\lambda$, create $\theta$ for the surrogate model.

\begin{eqnarray}
\Phi&=&([\mathbf{x}_{1,s},\ldots,\mathbf{x}_{1,s+\psi}], \ldots, [ \mathbf{x}_{M,s},\ldots,\mathbf{x}_{M,s+\psi}])\nonumber\\
\lambda &=&([F_{1,s},\ldots,F_{1,s+\psi}], \ldots,  [F_{M,s},\ldots,F_{M,s+\psi}])\nonumber \\
\Theta &=& [\Phi, \lambda]
\label{data}
\end{eqnarray}

where $\bf{x}_{i,s}  $  represents the given particle  from the swarm, $s$, $F_{i,s}$ is the output from the true fitness, and $M$ is the   number of swarms. Therefore, the   surrogate training dataset ($\Theta = [\Phi, \lambda]$) is made up of   input features ($\Phi$) and response ($ \lambda$) for  the particles that get collected in  each surrogate interval  ($s+\psi$). The  pseudo-fitness is given by $\hat{y} = \hat{f}(\Theta)$.

\subsubsection{Surrogate-assisted Framework}

In \textcolor{black}{Algorithm 1}, we present further details about  our framework that features surrogate-assisted optimisation using distributed swarms. 

We implement the algorithm using distributed computation over CPU cores, as shown in \textcolor{black}{Algorithm 1}. The manager process is shown in black where inter-process communication among swarms takes place which  exchange parameters at regular intervals (given by $\phi$). Furthermore, the surrogate model is also trained at regular intervals ($\psi$). The parallel swarms of the distributed framework have been highlighted in pink in \textcolor{black}{Algorithm 1}.
Stage 0 features the initialization of particles in the swarm. We begin the optimisation process by initialising all the swarms (Stage 0.1) in the ensemble with random real numbers in a range as required by the optimisation function (model).
 
Once the swarms are initialised, we begin the evolution (optimisation) by first evaluating the particles in the swarms using the fitness (objective) function. \textcolor{black}{Hence, we iterate over surrogate interval ($\psi$)  and evolve each swarm for $\phi$ generations, both user-defined parameters.} We update the best particle and best fitness for each of the respective swarms in the ensemble afterwards. Once these basic operations are done, we begin the evolution   where we create a new set of swarms for the next generation by velocity and position update (Stage 1.1). 
 
The crux of the framework is when we consider whether to evaluate the fitness function (true fitness) or to use the surrogate model of the fitness function (pseudo-fitness) when computing fitness values. Stage 1.2 shows how to update the fitness using either surrogate or true fitness of the particle depending on the interval and $S_{prob}$. Initially, this is not done until the very first surrogate interval is not reached, where all the fitness evaluations are from the true function. In Stage 1.3, we calculate the moving average of the past three fitness values for a particular particle by  $F_{past}$ = mean($F_{g-1}, F_{g-2}, F_{g-3}$) to  combine with the surrogate model prediction (Stage 1.4). \textcolor{black}{This is done so that we incorporate the recent history of the true model fitness, with the surrogate-based estimated  fitness, which is motivated by the autoregressive moving average (ARMA) model. Note that if present, the  surrogate estimation fitness will be also considered as part of  the past three fitness values. In the case when surrogate fitness is included, we note that additional errors will be added; however, this will be further averaged with  true values as there cannot be two surrogate fitness values.  } In Stages 1.5 and 1.6, we calculate actual fitness and save the values for future surrogate training. Our swarm particle update depends on $x_{pbest}^i$ and $x_{gbest}^i$, and due to poor surrogate performance, some of the weaker particles can get higher fitness scores. In order to avoid this issue, we ensure that $x_{pbest}^i$ and $x_{gbest}^i$ are from true fitness evaluation. In Stage 2.1, given a regular interval ($\phi$ generations), we  prepare an exchange of selected particles with neighbouring swarms via elitism, where we  replace a given percentage of weak particles (given by fitness values). \textcolor{black}{ $\beta$ is a user-defined parameter that determines how much of the swarm particles need to be exchanged. We have given recommendations for $\beta$ value in the design of experiments. We select only 20\% of elitist values as we do not want the majority of the swarms to be similar and maintain diversity. }

We note that before we consider the use of pseudo-fitness, we need to train the surrogate model with the same training data which is created from the true fitness. Hence, we need to collect the training data for the surrogate model from all the swarms in the ensemble. In Stage 3.0, the algorithm  uses surrogate training data collected from Stage 1.6  ($\Theta$ as shown in Equation \ref{data}). In Stage 4, the algorithm trains the surrogate model in the manager process with data from Stage 3.  The knowledge from  the trained surrogate model is then used in the fitness estimation as shown in  Stage 1.4. Stage 5.0 implements  the termination condition where  the algorithm signals the manager process to  decrement  the number of swarms alive (active) in order to terminate the swarm process. This is done  when the maximum number of fitness evaluations has been reached  ($T_{max}$) for the particular swarm. We use  a neural network-based surrogate model with Adam optimisation \citep{Adams2017}. In order to validate the performance of the algorithm, we measure the quality of the surrogate estimate using the root mean squared error (RMSE):

\begin{equation}
 RMSE = \sqrt{\frac{1}{N} \sum_{i=1}^{N} (F_i - \hat{F_i})^2}
 \end{equation}
 
 \noindent where $F_i$ and $\hat{F_i}$ are the true and pseudo-fitness values, respectively. $N$ denotes the number of times the surrogate model is used for estimation. \textcolor{black}{Figure \ref{fig:surrogate} provides a visual description of the proposed algorithm where multiple swarms are  executed using the manager process that also controls  the particle exchange and surrogate model update.}

\begin{figure*}[htp!]
\includegraphics[width=\textwidth]{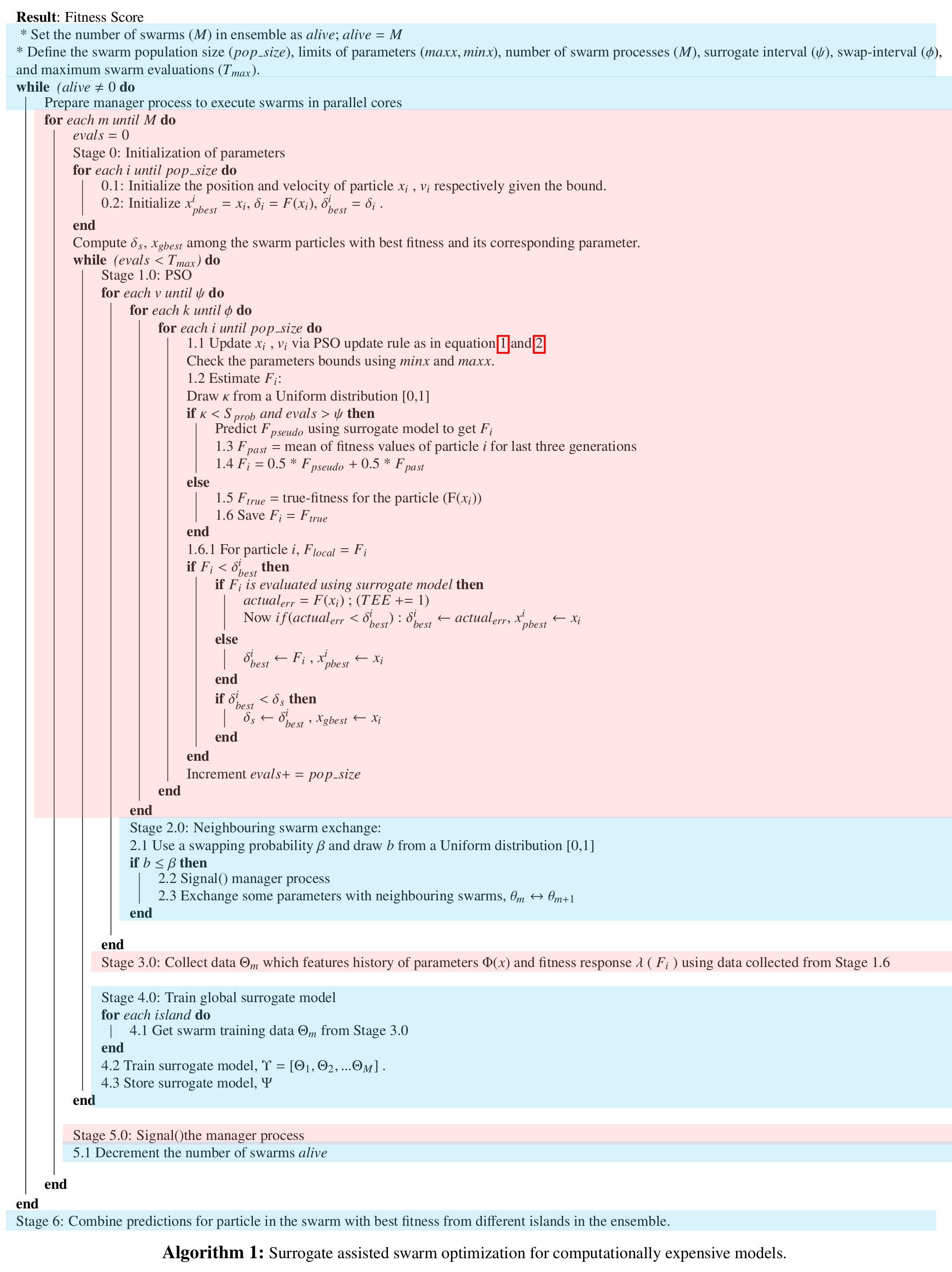}
 \label{fig:ptfnn}
\end{figure*}

\newpage

\begin{figure*}[htp!]
\includegraphics[width=\textwidth]{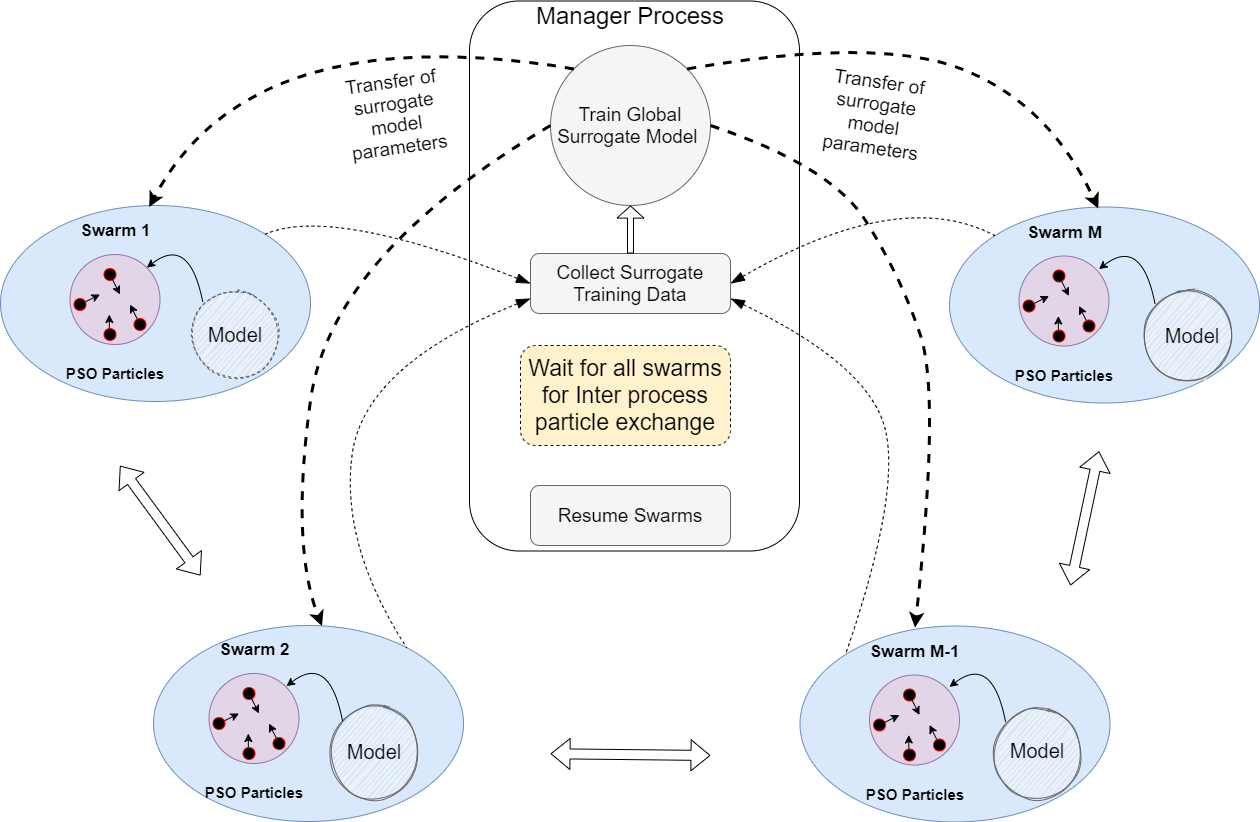}
 \caption{Surrogate-assisted distributed swarm optimisation  features surrogates to estimate the   fitness of expensive models or functions.   }
 \label{fig:surrogate}
\end{figure*}

 \section{Application: Landscape evolution models}

 \begin{figure}[htbp!]
  \begin{center}  \begin{tabular}{cc} 
 
 \subfigure[Continental-Margin]{\includegraphics[width=90mm]{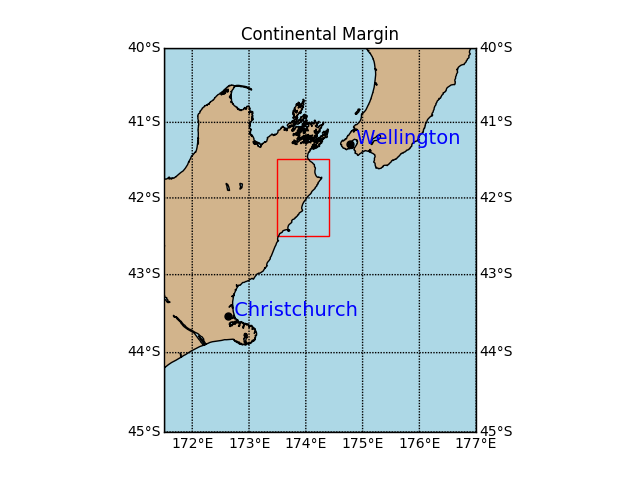}}

    \end{tabular}
    \caption{Location of (a) Continental-Margin problem shown taken from South Island of New Zealand 
     }
 \label{fig:cm-map}
  \end{center}
\end{figure}

  Similar to optimizing mathematical functions, in certain problems,  we are required to evaluate a score using some simulation or  computationally expensive process, such as geoscientific models \citep{
martin2004numerical,unger2012visual,chen2014landscape}. Landscape evolution models (LEMs) are a class of geoscientific  models that evolve a given topography over a given time with given geological and climate conditions such as rock erodibility and precipitation \citep{salles2018pybadlands}.  LEMs are used to model and understand the landscape and basin evolution back in time over millions of years showing surface processes such as the formation of river systems and erosion/deposition, where there is  movement of sediments from source (mountains) to sink (basins) \citep{chen2014landscape}. LEMs help geologists and paleoclimate scientists understand the evolution of the planet and climate history over millions to billions of years; however, there are major challenges when it comes to data. LEMs generally require data regarding paleoclimate processes which is typically unavailable; and hence, we need to estimate them with   methods such as  MCMC sampling  \citep{chandra2018PT-Bayes_,Chandra2018_Bayeslands_}. \textcolor{black}{There is limited work in the literature where optimisation methods have been used to estimate the unknown parameters for LEMs, which is the focus of this study. } We note that typically, LEMs are computationally very expensive which is dependent on the resolution of the study area (points/kilometer) and the duration of evolution (simulation) back in time (millions of years). Hence, large scale study areas  can take from hours to days to run a single LEM even with  parallel computing \citep{salles2018pybadlands}. There is   no gradient information in the case of Badlands LEM used for this study,  and hence estimating the  model parameters is a challenge.

\begin{figure}[htbp!]
  \begin{center}
    \begin{tabular}{cc}  

      \subfigure[CM  initial 
topography]{\includegraphics[width=65mm]{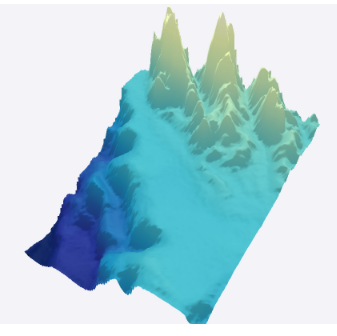}} \\
 

      \subfigure[CM   ground-truth 
topography]{\includegraphics[width=65mm]{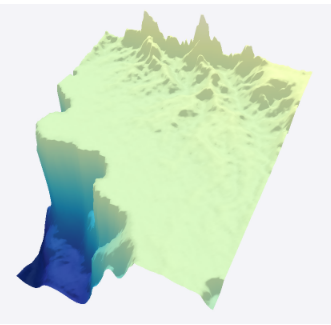}}\\


      \subfigure[\textcolor{black}{CM erosion/deposition map where yellow dots show the location (well) for computing sediment fitness (Equation 6). } ]{\includegraphics[width=80mm]{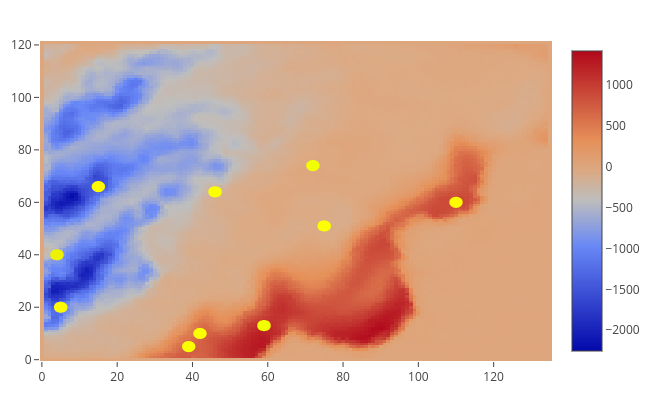}}\\

    \end{tabular}
    \caption{ 
The figures show the initial and evolved ground-truth topography and  erosion/deposition   after one million years for CM  problems taken from \citep{chandra2018PT-Bayes_,Chandra2018_Bayeslands_}.   The x-axis denotes the latitude and the y-axis denotes the longitude from the location given in \textcolor{black}{Figure  \ref{fig:cm-map}}, and the elevation is given in meters  which is further shown as a colour bar.     }
 \label{fig:craterdata}
  \end{center}
\end{figure}

  In order to demonstrate the optimisation procedure, we select problems where synthetic initial topography has been created using present day topography  and used in our previous work \citep{chandra2018PT-Bayes_,Chandra2018_Bayeslands_}.  
  The selected LEM features a   continental margin (CM) problem that is selected taking into account computational \textcolor{black}{time of  a single model run}  as it takes less than three seconds on a single central processing unit (CPU).  The CM problem initial topography is selected from the present-day South Island of New Zealand as shown in   Figure \ref{fig:cm-map} which covers 136 by 123 kilometres.   
 We provide a visualization of  the  initial and final topographies along with an erosion/deposition  map for \textcolor{black}{CM problem 
in  Figure} \ref{fig:craterdata}. The CM features six free parameters (Table  
\ref{tab:truevalues}). The notable feature of  all three problems is that they model both elevation and erosion/deposition topography.  We use the  initial topography (Figure \ref{fig:craterdata}) and the true values given in Table \ref{tab:truevalues}, and run the Badlands LEM by simulating 1 million years to synthetically generate the ground-truth topography. We then create a fitness function with the ground-truth topography and set experiments so that the proposed  optimisation methods can get back the true values.   The details about the fitness function are given in  the following section.

\begin{table*}[htbp!]
\small 
\centering
 \begin{tabular}{ l c c c c  c  c c } 
 \hline 
 \hline
  Issue  & Rainfall (m/a)&  Erod. &  n-value   & m-value & c-marine & c-surface & Uplift (mm/a)  \\  
 \hline

True-values  & 1.5  & 5.0-e06  & 1.0  & 0.5 &  0.5  & 0.8 & - \\  

Limits & [0,3.0 ]  & [3.0-e06, 7.0-e06] & [0, 2.0]  & [0, 2.0] & [0.3, 0.7]  & 
[0.6, 1.0] & - \\  
 \hline

 \end{tabular}
 
\caption{True values and limits of parameters. }
 \label{tab:truevalues} 
\end{table*}

\subsection {Fitness function}

 The Badlands LEM produces a simulation of  successive time-dependent topographies; however,  only the final topography $\mb D_T$ is used for topography fitness since no successive ground-truth data  is available. The sedimentation (erosion/deposition) data  is typically used to ground-truth the time-dependent evolution of surface process models that include sediment transportation and deposition \citep{salles2018pybadlands,chandra2019bayeslands}.

  We adapt the fitness function from the likelihood function used in our previous work that used Bayesian inference via MCMC sampling  for parameter estimation in Badlands LEM \citep{chandra2018PT-Bayes_}.  $\Omega$  represents the vector of free parameters, such as precipitation rate and erodibility which are independent and optimised by our proposed algorithms based on PSO.  The initial topography is given as  a two-dimensional matrix  $\boldsymbol D_{u,v}$, where corresponds to the location which is given by the latitude  $u$  and longitude  $v$ (Figure 2). 
  Hence, our topography fitness function  $F_{topo}$  for  the topography is  given by computing $f(.)$ that represents the final topology (at final time $t=T$) by Badlands LEM. 
  
\begin{equation}
 F_{topo}(\Omega) =  \sqrt{1/N \sum_{i=0}^N \left(\bar{D}_{i}-f_{i}(\Omega)\right)^2}
\end{equation}
where, \textcolor{black}{ $\nu$ is the number of observations.}  

Badlands LEM produces a sediment erosion/deposition topography at each time frame. We use a selected vector of locations  (Figure 3 - Panel c)   at time  ($\mb z_t$)  simulated (predicted) by the Badlands   LEM   for  given  
$\Omega$. The sediment fitness  $F_{sed}(\Omega)$ is given below
\begin{equation}
 F_{sed}(\Omega) =   \sqrt{1/(T+J) \sum_{t=1}^T\sum_{j=0}^J \left(z_{j,t}-g_{j,t}(\Omega)  \right)^ 2 }
\end{equation}

\textcolor{black}{The total fitness is the combination of the topography and sediment fitness. Note that the initial topography should not be confused with the initial state of the swarms of the PSO. The initial topography is simulated using pre-day topography of the region and hence not generated via any distribution.}

 \section{Experiments and Results}
 
\textcolor{black}{In this section, we present the results  of our framework on synthetic benchmark functions and    Badlands LEM. The  experiments  consider  a  wide  range  of performance measures which includes optimization performance in terms of fitness score, computational time, and accuracy of the  surrogate model.
}
  
\subsection{Experiment design}

We provide the experimental design and parameter setting for our experiment as follows. We implement distributed swarm optimization using parallel computing and inter-process communication where the swarms can have separate processes and exchange solutions (particles) with Python multi-processing library \footnote{\url{https://github.com/sydney-machine-learning/surrogate-assisted-distributed-swarms}}. 

The synthetic benchmark optimisation  functions are given in Equation 6 (Spherical), Equation 7 (Ackley), Equation 8 (Rastrigin), and Equation 9 (Rosenbrock). \textcolor{black}{In Equation 7, we use the user-defined parameters, $a = 20$ and $b = 0.2$. Similarly in Equation 8,  $a$ and $b$ and $c$ are user-defined parameters. } Note that these functions are chosen due to different levels of difficulty in optimisation and the nature of their fitness landscape.   The Spherical function is considered to be a relatively easier optimisation problem since it does not have interacting variables. \textcolor{black}{Ackley and Rastrigin are known to have many local minimums, while Rosenbrock is known as the valley-shaped function. }

\begin{equation} 
 {f(\mathbf{x}) = f(\mathbf{x}_1, \mathbf{x}_2, ..., \mathbf{x}_n) = {\sum_{i=1}^{n} \mathbf{x}_i^{2}}}
\end{equation}

 \begin{equation} 
f(\mathbf{x})=\sum_{i=1}^{n}[b (\mathbf{x}_{i+1} - \mathbf{x}_i^2)^ 2 + (a - \mathbf{x}_i)^2
\end{equation}

\begin{equation*} 
{f(\mathbf{x}) = -a * exp(-b\sqrt{\frac{1}{n}\sum_{i=1}^{n}\mathbf{x}_i^2})-exp(\frac{1}{n}\sum_{i=1}^{n}cos(c\mathbf{x}_i))+ a + exp(1)}
\end{equation*}

 \begin{equation} 
 {f(\mathbf{x})=10n + \sum_{i=1}^{n}(\mathbf{x}_i^2 - 10cos(2\pi \mathbf{x}_i))} 
\end{equation} 


We use ($M=8$) swarms which run as parallel processes (swarms) that \textcolor{black}{inter-communicate with each other at regular interval ($\psi=1$). We use the same values for certain hyper-parameters ($\psi=1$ and $\phi=1$) to simplify the algorithm. We exchange of a subset of particles (best 20\% from the population) with the neighbouring island (worst 20\% of the population). This is implemented  by setting $\beta=0.2$ in Algorithm 1.} We use the population size of 20 particles per swarm, the inertia weight $w=0.729$ with social and cognitive coefficients($c1=1.4$ and $c2=1.4$). We determined these values for parameters in the trial experiments by taking into account the performance on different problems with a fixed number of evaluations. We use minimum and maximum bounds on the parameters as described in Table 1, respectively. \textcolor{black}{We run experiments for 30 and 50 dimension  (30D and 50D) instances of the respective benchmark problems. The dimension of our problem for synthetic functions (eg. Rosenbrock) is much larger than LEM since in the literature, 30D optimization functions are more common. We set the total number of function evaluations  to 100,000 and 200,000, respectively. In the case of the Badlands model, we use 10,000 function (model) evaluations for the CM problem.}
    
We used the \textit{PyTorch machine learning library} \footnote{PyTorch: \url{https://pytorch.org/}} for implementing the surrogate model that uses a neural network model with Adam learning. The  surrogate neural network model architecture is given in Table \ref{tab:acq_arch},  where the input dimensions are defined, i.e. 30D and 50D instances of  the benchmark  functions. In a similar way, we extend our approach for optimization in the Badlands model with the CM problem, featuring a 6-dimensional  (6D) search space. The first and second hidden layers of the  neural network-based surrogate model are shown in Table    \ref{tab:acq_arch}.  We note that the output layer in the model for all the problems contains only one unit which provides the estimated fitness score.

 \begin{table}[htbp!]
\centering
\small
 \caption{ Model architecture for surrogate model}
\label{tab:acq_arch}
\begin{tabular}{lllll}
\hline
 \hline
Problem& Hidden($h_1$)& Hidden($h_2$)&\\
\hline
\hline
Rosenbrock 30D& 30 & 15 \\
Rastrigin 30D& 30 & 15 \\
Ackley 30D& 30 & 15   \\
Spherical 30D& 30 & 15 \\
Rosenbrock 50D& 50 & 25 \\
Rastrigin 50D& 50 & 25 \\
Ackley 50D& 50 & 25   \\
Spherical 50D& 50 & 25 \\
Badlands& 20 & 10 \\
\hline
\end{tabular}
\end{table}

\subsection{Results for synthetic benchmark functions}

We first present the results  (fitness score) for different problems using  serial (canonical) PSO, distributed PSO (D-PSO) and surrogate-assisted distributed swarm optimisation (SD-PSO)   as shown in Table \ref{tab:opt-syn}. We use a fixed surrogate probability \textcolor{black}{($s_{prob}=0.5$)} and  present results featuring mean, standard deviation (std), and best and worst performance for   over 30 independent experimental runs with different random initialisation in swarms as shown in Table  \ref{tab:opt-syn}. Note that lower fitness scores provide better performance. 
 
We find a significant reduction in elapsed time for all the problems in Table \ref{tab:opt-syn}. Note that we added a  0.05 seconds time  delay   to the respective problems in order to make them slightly computationally expensive to depict real-world application problems (models).  When we consider serial PSO with D-PSO and SD-PSO in terms of optimisation performance given by the fitness score, we find that D-PSO improves the PSO significantly for 30D and 50D cases of Rosenbrock, Spherical and Rastrigin problems. We also see improvement in Ackley's problem, but it's not as large as in previous problems. We find the variation in the results given by the  standard deviation is lowered highly  with D-PSO which shows it is more robust to initialization and has the ability to  provide a more definitive solution.

\textcolor{black}{Moreover, in comparison of  SD-PSO with D-PSO,   we mostly get better fitness with SD-PSO, with a reduction of computation time due to the use of surrogates.}  Despite the use of surrogate-based fitness estimation, we observe that the fitness score has not greatly  depreciated. \textcolor{black}{It is interesting to note in Ackley 30 and 50D case, addition of surrogates  improved the fitness score.} We find a 30\% reduction in computational time, it is expected this to increase if we had more time delay (rather than 0.05 seconds), which  will be shown in the Badlands LEM  experiments to follow.

The RMSE of prediction of fitness by the surrogate model is shown in Table \ref{tab:surr_acc}. We notice large RMSE values for the Rosenbrock problem  when compared to the rest.   In  surrogate prediction accuracy (RMSE) given in Figure  \ref{fig:Surrogate training}, we observe  a constant reduction in RMSE with intervals (along the x-axis) in   Ackley (30D and 50D), and Spherical (30D) model functions. In other problems, the RMSE is lower    towards the end, but the trend is not that smooth. We note that in the Rosenbrock function, there exists an interval where our surrogate performs poorly causing a major decrease in accuracy.                                          

\begin{figure*}[htbp!]
  \begin{center}
    \begin{tabular}{cc} 
    
    \subfigure[\textcolor{black}{Fitness score}]{\includegraphics[width=75mm]{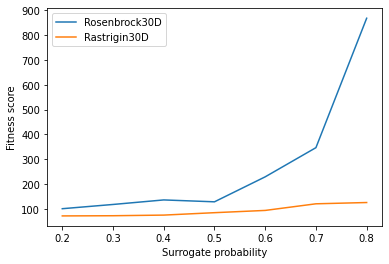}} 

      \subfigure[\textcolor{black}{Computational time} ]{\includegraphics[width=75mm]{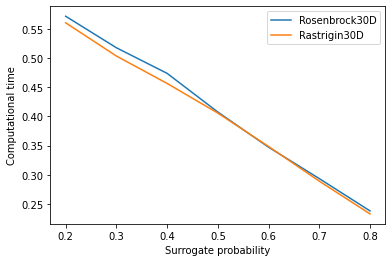}} \\
       
    \end{tabular}
    \caption{ 
The figure shows the effect of surrogate probability on fitness score and computational time. }
 \label{fig:surr_prob}
  \end{center}
\end{figure*}

\begin{figure*}[htbp!]
  \begin{center}
    \begin{tabular}{cc} 
    
    \subfigure[Ackley]{\includegraphics[width=75mm]{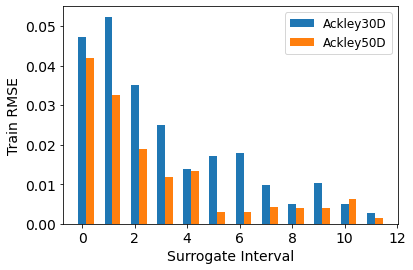}} 

      \subfigure[Spherical]{\includegraphics[width=75mm]{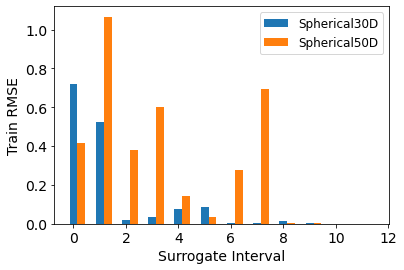}} \\
      
       \subfigure[Rastrigin]{\includegraphics[width=75mm]{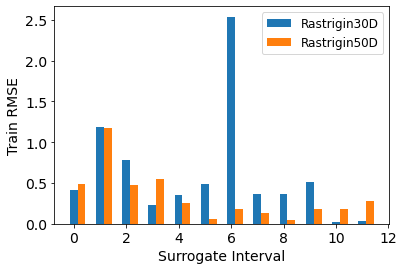}} 
       
       \subfigure[Rosenbrock]{\includegraphics[width=75mm]{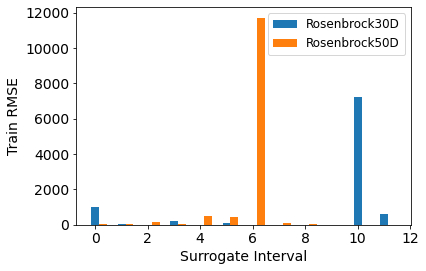}} \\
       
       \subfigure[   Badlands model]{\includegraphics[width=75mm]{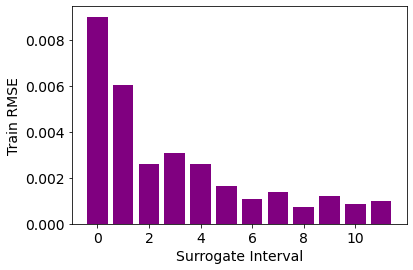}}
 
    \end{tabular}
    \caption{ RMSE of surrogate training during different surrogate intervals.  }
 \label{fig:Surrogate training}
  \end{center}
\end{figure*}

\begin{figure*}[htbp!]
  \begin{center}
    \begin{tabular}{cc}  
 \subfigure[Rosenbrock 30D 
]{\includegraphics[width=80mm]{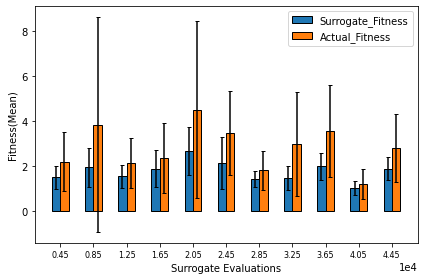}} 

      \subfigure[Rosenbrock 50D
]{\includegraphics[width=80mm]{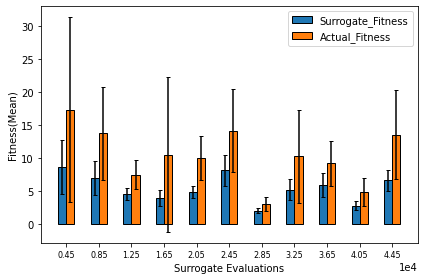}} \\
 
\subfigure[Spherical 30D
]{\includegraphics[width=80mm]{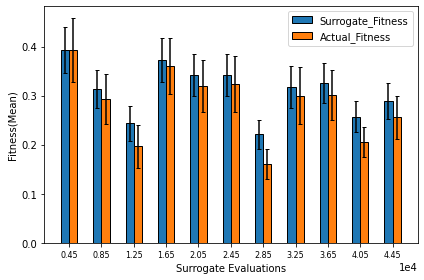}}

      \subfigure[Spherical 50
]{\includegraphics[width=80mm]{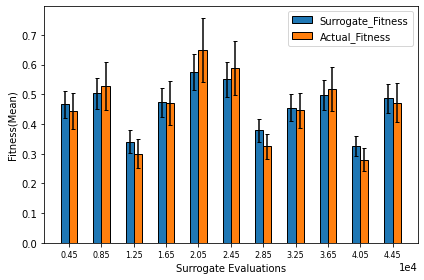}}\\

    \end{tabular}
    \caption{ 
Surrogate fitness versus actual fitness score over number of fitness evaluations (surrogate evaluations) for different problems.  }
 \label{fig:surr_hist0}
  \end{center}
\end{figure*}

\begin{figure*}[htbp!]
  \begin{center}
    \begin{tabular}{cc}  
 
\subfigure[Rastrigin 30D
]{\includegraphics[width=80mm]{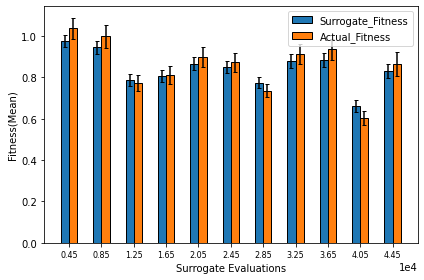}}

      \subfigure[Rastrigin 50D ]{\includegraphics[width=80mm]{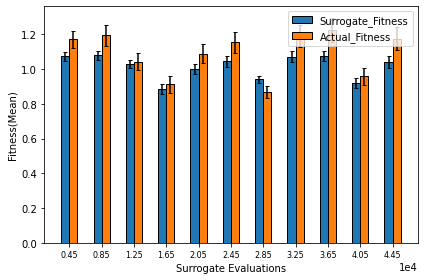}}\\
      
\subfigure[Ackley 30D
]{\includegraphics[width=80mm]{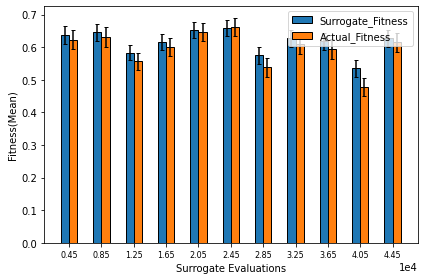}}

      \subfigure[Ackley 50D ]{\includegraphics[width=80mm]{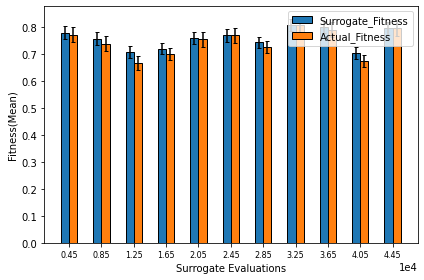}}\\

    \end{tabular}
    \caption{ Surrogate fitness versus actual fitness score over number of fitness evaluations (surrogate evaluations) for different problems.    }
 \label{fig:surr_hist1}
  \end{center}
\end{figure*}

 Figure \ref{fig:surr_hist0} and \ref{fig:surr_hist1} provide a visual description of surrogates prediction quality in the evolution process. We show the bar plots for mean values with a 95\% confidence interval (shown by error bars) of the actual fitness and pseudo-fitness at regular intervals for different benchmark problems.  We notice that the Rastrigin and Ackley problems have a better surrogate prediction with better confidence intervals when compared to the Spherical problem.                                                                            Finally, we evaluate the effect of the surrogate probability for Rosenbrock and Rastrigin 30D problems. Figure \ref{fig:surr_prob} (Panel a) provides a graphical analysis of the effect of the fitness score and computational time    (Panel b)  given different surrogate probabilities. We observe that the computational time decreases linearly with an increase in the surrogate probability. \textcolor{black}{On the other hand, the  fitness score degrades with an increase of  surrogate probability (Rosenbrock 30D); however,  with an elbow-shaped curve -- a trade-off can exist between time and optimization performance   (surrogate probability of 0.6).  \textcolor{black}{In the case of the Rastrigin 30D, there is not a large difference in loss of fitness (surrogate probability $\leq$ 0.5) when compared to the Rosenbrock 30D; we note that these problems have distinctly different fitness landscapes which could explain about the difference in the performance. }}

\begin{table*}[htbp!]
\centering
\small
 \caption{Optimization results on benchmark functions}
\label{tab:opt-syn}
\begin{tabular}{ c c c c }
\hline
 \hline
Problem&Method&Fitness Score& Elapsed Time \\

& &[mean\quad   std\quad    best\quad   worst] &  (minutes) \\
\hline
\hline
 Rosenbrock 30D&PSO&183.95\quad 109.94\quad 60.88\quad 455.01& 84.39\\
&D-PSO&108.78\quad 38.0\quad 55.0\quad 178.0& 11.22\\ 
 &SD-PSO (0.5) &149.7\quad 50.91\quad 67.92\quad 270.18& 7.113  \\
\hline

Rosenbrock 50D&PSO&1285.41\quad 559.54\quad 432.89\quad 2519.02& 84.37\\
&D-PSO &1269.03\quad 390.0\quad 78.0\quad 2084.0& 11.24\\
&SD-PSO(0.50) &2449.32\quad 1093.8\quad 685.48\quad 5298.79 & 7.114  \\

\hline
Spherical 30D &PSO&9.06\quad 9.22\quad 1.43\quad 37.85 & 84.35\\  
&D-PSO& 4.79\quad 3.0\quad 1.0\quad 12.0& 11.31  \\

&SD-PSO&1.82\quad 0.74\quad 0.9\quad 3.33  & 7.112 \\
\hline
Spherical 50D&PSO& 252.96\quad 102.69\quad 67.75\quad 426.74& 84.38 \\
 &D-PSO &175.2\quad 45.0\quad 118.0\quad 256.0& 11.23  \\
&SD-PSO(0.50) & 111.85\quad 48.36\quad 34.05\quad 239.828& 7.113 \\
\hline
Rastrigin 30D&PSO  &72.33\quad 19.72\quad 27.05\quad 112.21 & 84.38\\
 &D-PSO & 62.61\quad 12.0\quad 41.0\quad 90.0 & 11.20 \\ 
&SD-PSO(0.50) &77.23\quad 12.57\quad 52.96\quad 107.26& 7.113  \\
\hline
Rastrigin 50D&PSO  &188.34\quad 33.83\quad 130.96\quad 257.53& 84.39 \\
 &D-PSO &180.7\quad 24.0\quad 122.0\quad 215.0 & 11.1 \\ 
&SD-PSO(0.50) & 243.73\quad 25.1\quad 192.2\quad 307.82 & 7.114\\
\hline
Ackley 30D&PSO  &2.32\quad 0.39\quad 1.66\quad 3.03& 84.37\\
 &D-PSO & 1.83\quad 0.0\quad 1.0\quad 2.0& 11.21 \\ 
&SD-PSO(0.50) & 1.66\quad 0.33\quad 0.99\quad 2.24& 7.112  \\
\hline
Ackley 50D&PSO  &3.8\quad 0.38\quad 3.24\quad 4.58& 84.38 \\
 &D-PSO &3.42\quad 0.0\quad 3.0\quad 4.0 & 11.21 \\ 
&SD-PSO(0.50) & 3.34\quad 0.23\quad 2.76\quad 3.73& 7.113\\

\hline
\hline
\end{tabular}
\end{table*}

\begin{table*}[h]
\centering
\small
 \caption{Surrogate Accuracy }
\label{tab:surr_acc}
\begin{tabular}{c c c}
\hline

Problem &Surrogate Prediction RMSE &Surrogate Training RMSE  \\

 &  & [mean\quad std] \\
\hline
\hline

Rosenbrock 30D  & 368.33 & 7.68e+02\quad 1.97e+03   \\ 

Rosenbrock 50D & 284.73 & 1.09e+03\quad 3.45e+03 \\

Spherical 30D & 0.44 & 1.20e-01\quad 2e-01 \\ 
 
Spherical 50D  & 0.61 &7.10e-01\quad 8.75e-01 \\ 
  
Rastrigin 30D  & 0.46 &2.88e-01\quad 2.54e-01\\ 
Rastrigin 50D&  0.54 &7.33e-01\quad 1.35e+00 \\  

Ackley 30D  & 0.08 & 1.15e-02\quad 4.53e-03\\ 
Ackley 50D&  0.08 & 1.25e-02\quad 4.17e-03 \\  

\hline
Badlands (CM) & 0.0218 & 2.59e-3\quad 1.95e-3\\

\hline
\hline

\end{tabular}
\end{table*}

\subsection{Results for the Badlands model}

 Finally, we present results for  the case of the Badlands CM problem  which is a 6D problem. The results for CM problem highlighting our methods (PSO, D-PSO and S-DPSO) when compared to previous approaches (PT-Bayeslands and SAPT-Bayeslands)   are shown in Table \ref{tab:badlands-opt}. The results show the computational time and prediction performance of the Badlands model in terms of elevation and erosion/deposition RMSE (using Equations 4 and 5, respectively) given the optimised parameters. The results show the mean and standard deviation from 30 experimental runs from independent initial positions. We see a major reduction in computational time using  \textcolor{black}{D-PSO}  when  compared to PSO and find  consistent performance in terms of elevation and erosion RMSE. The experiments used  a surrogate model at an interval of 10 generations with  a probability of 0.5. We observe that the    \textcolor{black}{SD-PSO further improved the performance by reducing computational time using surrogates.}                                                                        The RMSE of the estimation of   fitness function by the surrogate model when compared to the actual Badlands model is shown in Table \ref{tab:surr_acc}. We note that the RMSE \textcolor{black}{in this case cannot} be compared to the synthetic fitness functions (eg. Rosenbrock) since the fitness function is completely different. In synthetic fitness functions, there is no data whereas in Badlands LEM, we use Badlands prediction and  ground-truth topography data to \textcolor{black}{compute}  the fitness.    Figure \ref{fig:Surrogate training} (Panel e) shows surrogate training accuracy (RMSE) for  different surrogate  intervals where we  observe  a constant improvement of performance by the surrogate model over time (surrogate intervals).  This implies that the surrogate model is improving as it gathers more data over time. 

\begin{figure*}[htbp!]
  \begin{center}
    \begin{tabular}{cc}  
 
\subfigure[Present day topography]{\includegraphics[width=75mm]{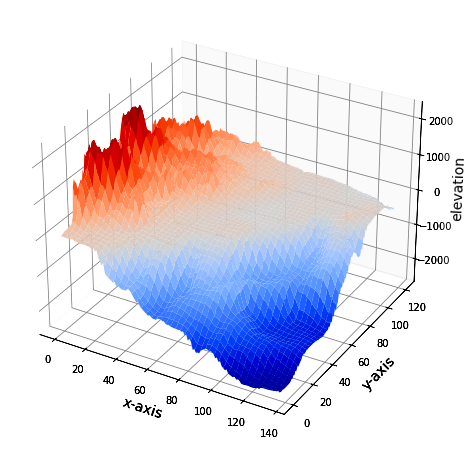}}

      \subfigure[200,000 years ]{\includegraphics[width=75mm]{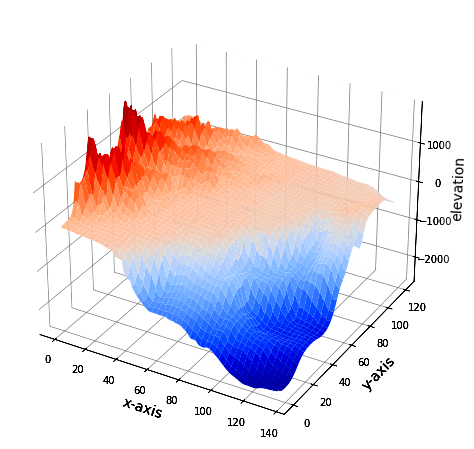}}\\
      
\subfigure[400,000 years]{\includegraphics[width=75mm]{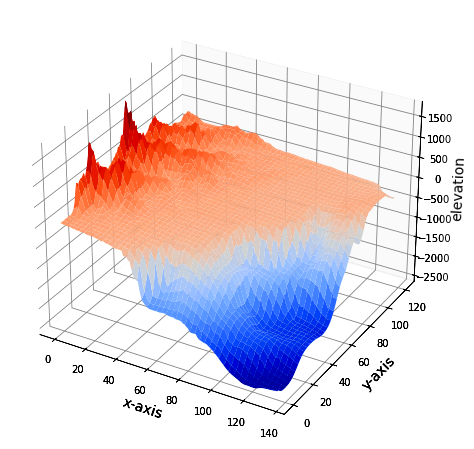}/}

      \subfigure[600,000 years]{\includegraphics[width=75mm]{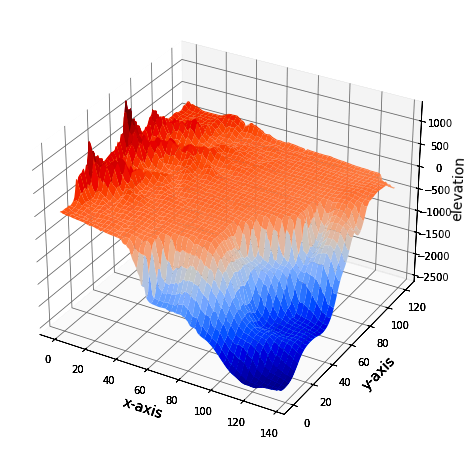}}\\
      
      \subfigure[800,000 years]{\includegraphics[width=75mm]{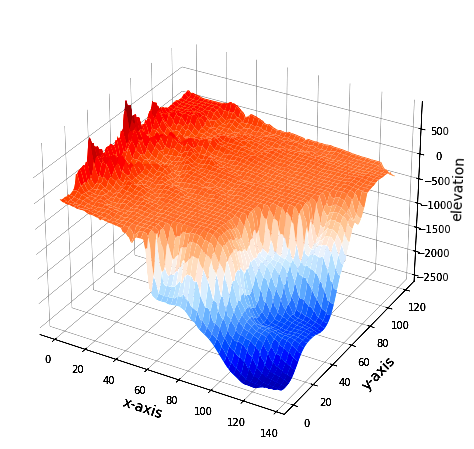}/}
      
    \subfigure[1 million years ]{\includegraphics[width=75mm]{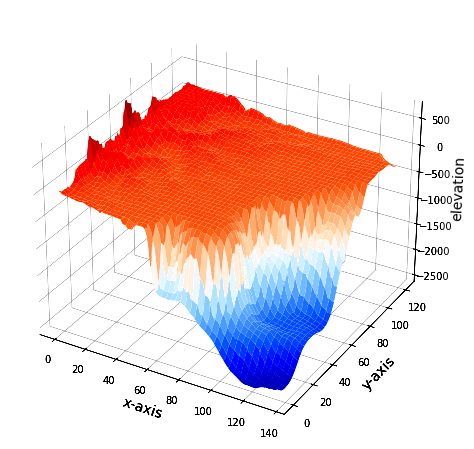}/}

    \end{tabular}
    \caption{This is how topology evolves according to our badlands model optimized using S-DPSO. Note that the distance over the x-axis and y-axis is given in kilometres (km) and the elevation is given in meters (m). }
 \label{fig:topology_opt}
  \end{center}
\end{figure*}

\begin{figure}[htbp!]
  \begin{center}
    \begin{tabular}{cc}  
    
      \subfigure[Cross-section reconstruction by optimizing Badlands CM model using S-DPSO. The ground truth represents  the actual topography, whereas Badlands prediction reports the mean topography predicted by   S-DPSO over 30 experimental runs. The $95\%$ confidence interval   represents the  uncertainty.]{\includegraphics[width=80mm]{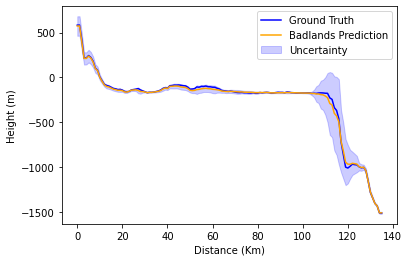}} \\

      \subfigure[Sediment prediction using S-DPSO. The ground truth represents the actual topography, whereas the Badlands prediction reports the mean topography predicted using S-DPSO over 30 experimental runs.]{\includegraphics[width=80mm]{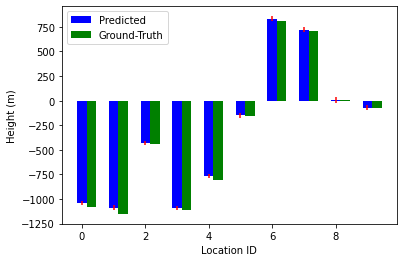}}

    \end{tabular}
     \caption{Prediction cross-section (Panel a) and sediment erosion/deposition (Panel b) with uncertainty given as 95 \% confidence interval from 30 experimental runs. }
 \label{fig:badlands_plots}
  \end{center}
\end{figure}

 In Figure \ref{fig:topology_opt}, we show the change in CM topology over selected time-slices  simulated by Badlands according to the parameters optimized by S-DPSO. The elevation RMSE in \textcolor{black}{Table} \ref{tab:badlands-opt} considers the difference between  ground-truth topography given in Figure \ref{fig:craterdata}. We notice that visually the final topography (present-day) in Figure    \ref{fig:topology_opt} (Panel f) resembles Figure \ref{fig:craterdata} (Panel b).  Furthermore, we show in Figure \ref{fig:badlands_plots} a cross-section (Panel a) for Badlands predicted elevation vs the ground-truth elevation for final or present-day  topography. We also show the bar plot (Panel b)  of predicted vs ground-truth sediment erosion/deposition at 10 selected  locations taken from Figure \ref{fig:craterdata} (Panel c). The cross-section and bar plots show that the Badlands prediction well resembled the ground-truth data, respectively. We obverse that the cross-section (Panel a) uncertainty is higher  for  certain locations as highlighted. The high uncertainty is in an area of a high slope below sea level which is reasonable given the effect of sediment flow due to precipitation.

 \begin{table*}[h]
\centering
\small
 \caption{Badlands performance }
\label{tab:badlands-opt}
\begin{tabular}{  c c c c c c }
\hline

\hline
\hline

 Method &Elevation  RMSE& Elevation  RMSE &Erosion  RMSE& Erosion  RMSE& Time  \\

  &  (mean)  & (std) &(mean) & (std)&  (seconds)\\
\hline
\hline

 PT-Bayeslands \citep{chandra2019bayeslands}, & 70.80  &10.03& 35.91 & 11.36 & 3243  \\ 

 SAPT-Bayeslands \citep{chandra2019PT-Bayeslands}   & 82.0 &8.23& 44.33 & 13.37 & 1859 \\
\hline
 PSO & 60.93 & 3.56 & 18.47 & 4.12 & 18480 \\

  D-PSO & 61.61 & 4.46 & 26.97& 5.39 & 4020 \\
 
  S-DPSO  & 61.17 & 4.57 &23.63 & 5.11 & 3462\\

\hline
\hline
\end{tabular}
\end{table*}

 \section{Discussion}

 \textcolor{black}{We note that the surrogate-assisted  method (SD-PSO) estimates the fitness and the velocity update in the next generation as done in a standard PSO (Equations 1 and 2). The surrogate fitness does not change the structure of the PSO, it is only used as a way to estimate the fitness and hence reduce the computational time. 
 We note that the performance of an optimisation method depends on the nature of the optimisation benchmark problems \cite{jamil2013literature} due to their fitness landscape modality, i.e., unimodal (Rosenbrock, Sphere) and multimodal (Rastrigin, Ackley), and  separability (Rosenbrock is considered  non-separable and Sphere separable). It is easier to construct optimisation methods for separable functions using a divide and conquer approach, which faces challenges in separable problems \citep{salomon1996re,chandra2016relationship}.  In terms of the results, we find that SD-PSO has done better than PSO and D-PSO (Table 3 and Table 5) for most problems. This could be done to the fitness estimation by the surrogate model, i.e. surrogate -based fitness estimations could have helped the algorithm in escaping local optima and hence it achieved better fitness.} In general, the results show that distributed surrogate-assisted swarm optimisation framework can   improve  performance by decreasing computation time while retaining optimisation accuracy (fitness).    

We highlight that the Badlands model does not provide gradient information regarding the parameters; and hence, only gradient-free optimisation and inference methods can be used. In our previous work,   MCMC sampling (PT-Bayeslands and SAPT-Bayeslands \citep{chandra2019bayeslands,chandra2019PT-Bayeslands}) has been used, where the parameter inference was implemented   via random-walk proposal distribution with MCMC replicas running in parallel. In this paper, the results show that the use of meta-heuristic (evolutionary) search operators from particle swarm optimisation provides   better search features. 
 The results motivate the use of the proposed methodology for  expansive optimisation models, which can feature other \textcolor{black}{geoscientific} models. Further use of surrogates in larger instances of the  Badlands  LEM can  provide a significant reduction in computational time.

  A major contribution of the framework is in the implementation using parallel computing, which takes into account inter-process communication when exchanging particles (solutions) during the optimisation process. In our proposed framework, the surrogate training was implemented in the manager process and the trained parameters were transferred to the parallel swarm processes, where the local surrogate model was used to estimate the fitness of the particle when required (Figure 1). This implementation  seamlessly updated the local surrogate model at regular intervals set by the user. We note that although less than eight parallel swarm processes were used, in large-scale problems, the same implementation can be extended and amended. We note that in the case when the number of parameters in the actual model   significantly increases, different ways of training the surrogate model can be explored. \textcolor{black}{We also note that the number of particles per swarm is strongly dependent upon the problem in hand \cite{kennedy1995particle,jain2022overview}.}
  
 Another major contribution from the optimisation process for the case of the landscape evolution model is the estimation  of the parameters, such as precipitation values in the Badlands model. \textcolor{black}{We note that MCMC sampling methods provide inference whereas, optimisation methods provide an estimation of parameters. The major difference is that we represent the parameters using probability distribution in the case of inference, whereas optimisation methods provide single-point estimates.} Through optimisation,  we can estimate what precipitation values gave rise to the evolution of landscape which resulted in the present-day landscape. The landscape evolution model hence provides a temporal topography map of the geological history of the region under study. These topographical maps, along with the optimised  values for geological and climate parameters (such as precipitation and erodibility) can be very useful to geologists and  paleoclimate scientists. 
   
  \textcolor{black}{Although PSO has been selected as the designated evolutionary optimisation algorithm, the framework has the flexibility to enable the implementation of other evolutionary algorithms depending on the application problem. A wide range of PSO variants exist in the literature with strengths and limitations \cite{ullmann2017comparison,wang2018particle,bonyadi2017particle}. In the case when the application  involves combinatorial optimisation or a scheduling problem, then an appropriate evolutionary algorithm would be needed. }

\textcolor{black}{Our experiments considered a simulation where the present-day topography of  a selected region was used as the initial topography. The Badlands model ran depicting a million years in time, to simulate successive topographies. However, in the area of LEMs, the focus  is generally  to understand the environmental and climate history back in time that led to present-day topography. In such cases, we need to estimate the initial topography (eg. a million years back in time), which can be based on present-day topography. The estimation of initial topography is an optimisation problem of its own and joint optimisation of parameters such as precipitation can make the process very complex. However, this can be addressed in future research.}

\section{Conclusions and Future Work}

We presented a surrogate  framework that features parallel swarm optimisation processes and seamlessly integrates surrogate training from the manager process to enable  surrogate fitness estimation. Our results indicate that the proposed surrogate-assisted optimisation method significantly reduces the computational time while retaining solution accuracy. In certain cases, it also helps in improving the solution accuracy by escaping from the local minimum via the  surrogates. Although we used PSO as the designated algorithm, other optimisation algorithms, such as genetic algorithms, evolution strategies, and differential evolution can also be used. 

In future work, the parallel optimisation process could be improved by  a combination of different optimisation  algorithms   which can provide different capabilities in terms of exploration and exploitation of the search space. The proposed framework can also incorporate other benchmark function models, particularly those that feature constraints and also be applied to discrete parameter optimisation problems which are expensive computationally.

 \section{Further Information}
 
 \subsection{Ethical approval}
 
 The data used  in the manuscript is openly available and does not need any ethical approvals. 

 \subsection{ Funding details}
 
 There are no external funding sources to report. 

 \subsection{ Conflict of interest}
 
 The authors do not have any conflict of interest with the manuscript and publication process. 
 
 \subsection{Availability of data and materials }
  We provide an open-source implementation of the proposed algorithm in Python along with data and sample results  \footnote{Surrogate-assisted distributed evolutionary algo: https://github.com/sydney-machine-learning/surrogate-assisted-distributed-evo-alg}. 
   
 \subsection{Authorship Contribution}
 
 R. Chandra contributed by writing, coding, experiments  and analysis of results. Y. Sharma contributed by coding, experiments, writing and analysis. 
  
\section*{Acknowledgement}

The authors would like to thank Ratneel Deo from the University of Sydney for  his support during the initial phase of this research.

\bibliographystyle{elsarticle-harv}

\bibliography{aicrg,2018,Chandra-Rohitash,Bays,sample,surrogate} 

\end{document}